# Media affordances of a mobile push-to-talk communication service

Allison Woodruff and Paul M. Aoki
Palo Alto Research Center, 3333 Coyote Hill Road, Palo Alto, CA 94304 USA
*woodruff@acm.org, aoki@acm.org*

**Abstract.** This paper presents an exploratory study of college-age students using two-way, push-to-talk cellular radios. We describe the observed and reported use of cellular radio by the participants, the activities and purposes for which they adopted it, and their responses. We then examine these empirical results using mediated communication theory. Cellular radios have a unique combination of affordances relative to other media used by this age group, including instant messaging (IM) and mobile phones; the results of our analysis do suggest explanations for some observed phenomena but also highlight the counter-intuitive nature of other phenomena. For example, although the radios have many important dissimilarities with IM from the viewpoint of mediated communication theory, the observed use patterns resembled those of IM to a surprising degree.

## Introduction

In the U.S., until recently, only government-licensed radio operators have been permitted to operate two-way radios that can communicate more than a few kilometers. In 1996, the deployment of digital cellular trunked-radio networks enabled wireless carriers to provide two-way radio services to consumers. Just as in wireless telephony, subscribers can communicate privately with other subscribers (as opposed to using a public channel) at distances that are limited only by their network's cellular coverage; unlike telephony, the service connects subscribers directly, without dialing delay.

One wireless carrier, Nextel Communications, provides mobile phones with conventional features such as voice telephony and voice mail; these same devices also support a two-way, push-to-talk service called Direct Connect™. This service is very popular, having 10 million subscribers and supporting nearly 50 billion Direct Connect™ calls in 2001, predominantly for business use (Nextel, 2002). Other U.S. carriers have announced plans to provide similar services in the very near future.

We report here on a qualitative study of the use of these two-way, push-to-talk cellular radio systems (henceforth *cellular radios*) by consumers. The study, intended to inform the design of a new mobile voice communication system, was conducted as lightweight, exploratory "fieldwork for design." Since our new system (described elsewhere (Aoki et al., 2003)) has the design goal of supporting out-of-workplace social relationships within gelled social groups, particularly those comprised of young adults, the study was aimed at collecting insights into emergent communication patterns, techniques and strategies developed by members of this target user population. At present, this population very rarely has access to cellular radios due to their cost. Accordingly, we provided college-age students with cellular radios for a week, observing their use of the devices and conducting interviews on an ongoing basis.

While there have been a few ethnographic studies of the use of short-range handheld radios, and the literature on the use of wireless telephony including text messaging (SMS) has grown tremendously in the last three years (e.g., (Harper, Green, & Brown, 2001; Katz & Aakhus, 2002)), we are not aware of any work that examines the use of cellular radio communication. From the perspective of mediated communication theory (Whittaker, 2003), cellular radios offer a unique combination of affordances – no other consumer service provides wide-area, private voice communication with a comparably lightweight interface.

In our study, we were intrigued to learn that the cellular radios supported a large number of activities – a range similar to that of IM, and wider than that of other technologies used by the participants. We discuss how different activities are related to the media affordances of cellular radios. Of particular interest is that, although the radios have many important dissimilarities with instant messaging (IM) from the viewpoint of mediated communication theory, the observed use patterns resembled those of IM to a surprising degree. For example, even though the radios completely lack the browseable history provided by textual message systems, we frequently saw the pattern of "intermittent conversation" associated with IM (Nardi, Whittaker, & Bradner, 2000), with its long pauses between turns.

The remainder of the paper is organized as follows. We first situate this study in prior work. We then provide details about the method used in the study, including details of the cellular radio service itself. Next, we describe specifics of the participants' use of cellular radios during the study. We then turn to a description of the radio service from the standpoint of mediated communication theory, applying Clark and Brennan's framework (Clark & Brennan, 1991) to a number of communication media including cellular radios.  Finally, we conclude with a discussion of our findings.

## Background and related work

Many of the phenomena seen in our data have been reported in studies of remote communication systems designed to facilitate "informal workplace communication" – the kind of "opportunistic, brief, context-rich and dyadic" (Nardi et al., 2000) interactions that happen between physically proximate workers (Whittaker, Frohlich, & Daly-Jones, 1994).  Such systems include classic open-channel environments such as video spaces (e.g., (Dourish, Adler, Bellotti, & Henderson, 1996)) and audio spaces (e.g., (Ackerman, Starr, Hindus, & Mainwaring, 1997)); more recently, there have been similar studies of various kinds of textual messaging tools such as IM (Isaacs, Walendowski, Whittaker, Schiano, & Kamm, 2002; Nardi et al., 2000).  A common finding is that providing a continuous connection between users facilitates a *bursty conversation* style, in which formal conversational openings and closings are infrequent even though long pauses occur between bursts of talk, as well as a more conventional *focused conversation* style.

Recent analyses of IM use have highlighted communication behaviors that are now strongly associated with IM.  Nardi *et al.* introduced the notions of *plausible deniability* (relying on the sender's lack of information, e.g., about the presence of the recipient, to excuse unresponsiveness) and *intermittent conversation* (interactions with long pauses between individual turns) (Nardi et al., 2000), and Voida *et al.* applied an affordance analysis to explain certain similarities between IM and voice communication (Voida, Newstetter, & Mynatt, 2002).  "Hanging out" – i.e., IM as social space – features prominently in teen and young adult use of IM (Grinter & Palen, 2002).

Another recent thread of research concerns the social use of mobile communication media, including SMS (Grinter & Eldridge, 2001).  Mobile phones, like IM, enable the construction of social spaces (Ito, 2001; Ling & Yttri, 2002) as well as affording other ritual communication (Taylor & Harper, 2002), though social uses must be carefully managed in public spaces (Palen, Salzman, & Youngs, 2001; Weilenmann & Larsson, 2001).  Mobile phones are widely used for *micro-coordination* – the use of dynamic, just-in-time activity coordination in lieu of extensive pre-planning (Ling & Yttri, 2002).

We are aware of very few studies of the use of handheld two-way radio ("walkie-talkie") communication.  The "Denver Project" introduced radios to Xerox service technicians, assessing their patterns of adoption as well as use (Orr, 1995).  A study at Interval Research supplied radios to teens for use during a weekend-long rock concert (Strub, 1997). To our knowledge, there have been no published studies of cellular radio, which affords wide-area communication.

# Method

In this section, we describe the procedure, participants, and equipment used.

## Procedure

The study took place in June 2002. Participants completed a pre-study questionnaire on demographics and on their use of communication technology. Participants were given cellular radios and asked to use them for approximately one week and provide feedback. They were given almost no explicit direction; specifically, they were rarely if ever encouraged to use the devices *per se*, and they were aware that the study was being conducted by employees of a company that neither manufactured cellular radios nor provided cellular radio service.

One author lived with four of the participants in their rental house during the majority of the study, and participated in many social activities during the week. This participating author also observed several of the participants at work, and conducted semi-structured interviews before, during, and after the study.

## Participants

We had two primary goals relating to the participants. First, we wanted the participants to be members of a pre-existing, gelled social group. Second, we wanted the participating author to be able to observe the participants throughout the day and night, in both public and private settings. We were able to accomplish both of these goals by working with a relative of the participating author. This relative was a participant in the study, and she selected and recruited all other participants by identifying the members of her own gelled social group. Because her relative is a trusted member of the social group, the participants gave the participating author privileged access to intimate details of their lives.

Participants were seven U.S. college undergraduates (five female and two male), all of whom were either 20 or 21 years of age and living away from their parents' homes. Most participants had known each other for several years and socialized frequently. Four of the participants (Erica, Julie, Ryan, and Todd) lived together in a rented house. Two additional participants (Kelly and Stephanie) rented an apartment together, and Dawn was a frequent visitor to that household. Julie and Todd were girlfriend and boyfriend. (Names have been changed to preserve the anonymity of the participants.)

In addition to social and leisure activities, all participants were enrolled in summer school, working, or both. Of relevance to future discussions is that

Dawn, Erica, Kelly, and Stephanie worked as waitresses and Todd had a computer science internship.

Participants very frequently used mobile phones: four of the participants had their own mobile phones, Julie and Todd shared a mobile phone, and Kelly did not have a mobile phone. Several participants frequently used IM, although others did not, e.g., one did not have a computer at home. Participant use of SMS, email, and home phones varied somewhat but was quite limited.

Equipment and system operation

We rented Motorola i1000 phones from a cellular service reseller. The phones were fairly large, measuring 114mm x 56mm x 30mm (4.5" x 2.2" x 1.2") and weighing 170g (6 oz.). Each participant, including the participating author, was given a phone and a single-earphone headset with a boom microphone; the phones could operate as a speakerphone (like a conventional handheld radio), as a telephone handset, or using the headsets. To control costs (which are extremely high for voice calls on rental phones), all features except cellular radio service were disabled, e.g., the phones could not be used to place telephone calls. No limits, time or otherwise, were placed on the use of the cellular radio service.

With this particular cellular radio service, one user uses a push-to-talk protocol to speak to another user. Specifically, if person A wishes to say something to person B, person A picks up their radio, selects person B's name from a menu and holds down a button. After a brief delay (variable, but generally about 1 sec.), a "go ahead" beep is heard and person A can speak. When person A releases the button, person A stops transmitting. After this, persons A and B simply push the button to speak (i.e., the menu is not used to select an addressee). Only one user can be speaking at a time; if person B pushes their button while person A is still speaking, the microphone will not be activated and the "go ahead" beep will not be heard until the channel is clear. After eight seconds have passed without a speaking turn, the radios reset and the menu can be used to select a new addressee; as an optimization, a "previous call" button reselects the last person spoken. Connections between individuals are called *private connections*.

A similar *group connection* mechanism can be used within predefined groups of cellular radio users. This mechanism is very similar to a channel in a handheld radio, except that the network provides access control – only pre-specified members are allowed to "tune in." The feature is of limited usefulness, since users can only "tune in" one group at a time, and (as with radio) they must already be "tuned in" to a given group to receive any of its messages.

When we say these connections are "lightweight," we usually refer to the fact that connecting is faster than speed-dialing or voice-dialing (particularly since there are no subsequent delays for dial-tone transmission or ringing).

# Observed and reported use

In this section, we discuss observed and reported use of the cellular radios. While these findings are of course specific to the participants in this study, this concrete view of *what actually happened* is helpful in that it grounds the more abstract analysis of media affordances in the following section. Naturally, our investigation revealed many of the concerns and issues seen in previous studies of communication technology, e.g., privacy and availability. Due to space constraints, we are unable to discuss all of these points at length. Instead, we focus on points that are most closely related to the particular combination of affordances offered by cellular radios.

## General patterns

In this subsection, we review the general use of the cellular radios.

*Overall availability.* Participants typically carried their cellular radios with them and kept them turned on. For example:

> Dawn: "I learned that I should just take it with me into the shower in case somebody's like trying to talk to me."

In fact, participants generally left their radios on at night and were sometimes awakened by them. (Interestingly, the ritual "good morning" message (Ackerman et al., 1997; Nardi et al., 2000) and "goodnight" message (Grinter & Eldridge, 2001; Taylor & Harper, 2002) usually observed with lightweight communication technology did not come up in observation or in interviews.)

*Overall level of activity.* While most participants were very strong adopters (we roughly estimate that participants used their cellular radios for on the order of tens of interactions per day), the level of activity was extremely variable. Sometimes there would be great bursts of activity, which were often interleaved with other social activity. For example, at a given moment in a car with five people, two participants in the car might be using their cellular radios in private connections with participants at other locations, while simultaneously participating in co-present conversations with people in the car. Such activity was often chaotic, especially since the cellular radios were typically used in speaker mode and since many conversations involved a lot of voice affect. At other times, participants would not use the cellular radios for long stretches, e.g., while they were engaged in an activity like watching a laser show.

*Dominant modes of use.* Participants used private connections much more frequently than group connections, partly because the group mechanism was cumbersome and partly to avoid annoying other participants.

Participants explained that speaker mode was preferable to headsets because the headsets were uncomfortable and because they drew attention from other people. Early in the week, the visual appearance of the headsets with the boom microphones was a source of amusement: several girls joked extensively that they were a girl band, and one boy dressed up in a security guard shirt that he happened to own. This type of affiliation has been reported elsewhere; recall that in (Strub, 1997), teens using short-range radios pretended to work for security.

Because the devices were used primarily in speaker mode, co-present individuals were able to overhear transmissions and consequently frequently became involved in interactions. Similar impact on co-present individuals has been observed for other media as well (Dourish et al., 1996; Palen et al., 2001; Weilenmann & Larsson, 2001). Neither transmitters nor receivers seemed particularly sensitive to the public nature of transmissions, although they did indicate some embarrassment about the fact that speech emanated from unexpected areas of their body, depending on where they had clipped their cellular radio. This sense of body parts such as thighs or hips "talking" did not result in a change in practice, i.e., they continued to use speaker audio.

## Comparison to patterns of use of other media

Participants said they used the cellular radios in preference to other technologies. In some cases, they used multiple communication media simultaneously; in these situations, the cellular radio had lower priority than media such as the telephone.

*Frequency of use.* Participants said that they spoke much more frequently on the cellular radios than on their mobile phones. This was consistent with our observations, participants' reports of cellular radio use, and the participants' responses to questions about frequency of mobile phone use in the pre-study questionnaire. While they were probably influenced by the fact that the cellular radio airtime was free, they reported that they were strongly influenced by the fact that the communication was so lightweight. Participants identified many specific situations in which they would use cellular radios in which they would not use mobile phones. For example, Erica reported that she would not use the phone to chat with people while she was at work, but she would use the cellular radio because she could start and stop the conversation quickly.

> Erica: "[I]f it weren't for the walkies talkies I just wouldn't talk to them."

Similarly, participants reported that they used the cellular radios in many situations in which they did not believe they would have used SMS (although most had fairly limited experience with it), saying that SMS was undesirable because of the effort of typing, the long delays, and the difficulty of communicating affect in textual communication.

People sometimes relayed or requested information over the cellular radios that would probably not have been worth sharing using more heavyweight mechanisms (analogous to effects reported for SMS relative to telephony (Ito, 2001)). For example, Erica said she would call people to ask questions which she felt would not be appropriate with the phone.

Erica: "It's really convenient with roommates. Cause you can ask em just stupid little questions, like, you know, 'Where's the extra toilet paper?' or something."

*Media use within-group vs. outside-of-group.* Cellular radios were almost always chosen in preference to mobile phones for communication within the group of study participants.

Dawn: "[T]hat would be like the drastic emergency thing. If I couldn't get anybody through this [the cellular radio], I would have to be like, okay, I have to use the phone now."

For at least one participant, they replaced the use of IM with other participants. Note that media-switching from cellular radios to other technologies was not observed or reported.

Cellular radios did not appear to have much impact on the use of communication technology or the frequency of communication with people who were not in the study. In direct terms, this is in part to be expected since almost none of the people they knew outside of the study owned cellular radios. However, in more indirect terms, one might expect a kind of "conservation of talk" – that intensified communication within the group might result in reduced communication with people outside of the group. This did not seem to occur.

*Precedence in cases of conflict.* Cellular radios were sometimes used at the same time as other communication media. Telephone calls, when they did occur, took precedence over the cellular radio. For example, if a participant were on the telephone and received a transmission on the cellular radio, they would typically ignore it.

Julie: "[A cellular radio transmission is] not like a phone call, so it's like a lower priority…"

Because of the prevalent use of speaker audio, it was not unusual for someone nearby to pick up the participant's cellular radio and reply on their behalf.

## Communicative activities and purposes

The cellular radios were used for a wide range of activities and purposes. While many of these have been reported elsewhere, the sheer diversity of uses for a single medium is interesting. We emphasize that all of these uses were originated by the participants. After we describe the activities, we introduce some key conversation patterns that occurred during many of the activities.

*Chit-chat*. The cellular radios were frequently used for brief, light conversation, particularly by the females. This often occurred as a "time filler" when participants were bored. For example, the girls who worked as waitresses would often use the cellular radios to talk with other people when they were bored at work. Casual conversation was also a popular activity when people were walking or driving somewhere by themselves.

Erica: "It's nice like walking back from work and stuff to be able to call people and just to chit-chat."

Dawn: "Kelly called me one time from her car, she's like, 'There's a fine guy in front of me, I'm following him!' I'm like, 'Go for it!'"

*Extended remote presence.* The cellular radios were often used for sharing extended activities or to allow participants to keep each other company for extended periods. We distinguish this from remote single-task participation, such as consultation during a trip to the grocery store. The difference is that what we call extended remote presence would continue for a very long period of time relative to a plausible phone call length; the periods could safely extend across local activities in which one participant or the other would have been likely to hang up a phone call. Julie "went on errands" with Kelly by speaking to her periodically in the cellular radio. While Erica and her boyfriend were at a baseball game, they checked in from time to time with participants who were eating dinner at a restaurant. The girlfriend/boyfriend pair kept in contact, speaking often while Todd was at work.

> Julie: "I talk to Todd a lot to just, you know, see what he's doing at work or just to bug him."

Another variation was interaction during sequences of short tasks. For example, Julie reported that she taught Ryan to make rice. She said the cellular radios were convenient for this; there was a lot of waiting while he completed steps in the recipe, so a phone call would have been inconvenient.

*Micro-coordination.* Cellular radios were often used for micro-coordination (Ling & Yttri, 2002). For example, one group went into a grocery store while another group went through a drive-through at a fast food restaurant; the group in the store spoke to the group in the drive-through about their order and then coordinated being picked up in the store parking lot.

*Substantive conversation.* The cellular radios were used for substantive conversations which were often multi-topical and sometimes lasted as long as thirty minutes, according to the participants. Such interactions often focused on the sharing of feelings and emotional support; for example, Erica contacted Kelly to tell her a story about a guy who was a "jerk" at work. The ability to communicate affect through voice was important.

> Julie: "I like it [the cellular radio] better than IM cause like in IM you always lose everyone's like um intonations when they say stuff…"

*Play.* The cellular radios were used for a number of playful activities. These largely relied on vocal "sound effects" (as in (Ackerman et al., 1997; Strub, 1997)) and the communication of affect, and so would not have been appropriate in a textual medium such as IM. They would also have been difficult to carry off if the participants would have had to call each other on the phone. For example, the two boys would frequently play military games like pretending to land airplanes, or spontaneously say things like "Damage report in sector 4" to each other. Other play included spontaneous singing of songs or repeated quoting of phrases from movies in funny voices.

Additional jokes relied on the fact that the cellular radios were used primarily in speaker audio mode, which created a space that drew in co-present parties (Dourish et al., 1996). For example, Dawn spoke to Julie's hamster without preface, saying, "Butterscotch, hello" at a time when she guessed that Julie might be near her hamster. In fact, pets were addressed and/or encouraged to "speak" during transmissions on multiple occasions.

*Communication patterns.* Several communication patterns occurred to a greater or lesser degree in each of the above activities. First, participants often had delayed responses to cellular radio transmissions. The participating author observed delays of approximately two to three minutes that occurred without apology or explanation. At other times, transmissions would simply go unanswered. As discussed further below, participants were well aware of their own behavior in this regard and accepted such behavior from others. Second, openings and closings would generally be omitted or reduced as compared with interactions in other media, such as phone conversations. Third, many other activities were routinely interleaved with use of the cellular radios. For example, participants might work on their computer or converse with co-present parties while periodically using their cellular radio to speak to other participant(s).

## Social impact

The cellular radios impacted the participants' social lives in a number of ways. Participants found that they spoke to each other more frequently and consequently had more awareness of each other's activities. This combination of more frequent conversation and increased awareness led to them seeing each other more often.

*Increased remote communication.* As discussed above, participants said they used the cellular radios more frequently than other technologies; as a direct consequence, they spoke to each other more often. Further, because it was so easy for participants to contact each other, expectations increased.

> Todd: "You know if you have it, people expect to be able to talk to you all the time."

Feelings on this increased availability were mixed. Some were pleased.

> Julie: "This is so great. I just lie here and all these people talk to me."

Others were more ambivalent. For example, while some participants liked being able to reach *other* people easily, those same participants sometimes found it irritating that other people were able to reach *them* easily.

Overall, increased availability was tolerable largely because there was a limited group of participants using the cellular radios, and therefore they were available only to close friends. They contrasted this with mobile phones, which they felt gave more people access to them.

> Julie: "[I]t's kind of fun too to have like this network where it's like only my friends and only like fun people are calling me. So I kind of like that."
> Kelly: "Yeah, I like that too."

When given hypothetical long-term use scenarios, participants were clear about the types of people with whom they would be willing to share cellular radio connectivity, and therefore increased availability. Participants were most interested in using cellular radios with roommates and close friends whom they saw regularly, they were less interested in using them with friends who lived far away, and they were generally *strongly* opposed to using with their families.

> Todd: "[I]f my dad had a radio-"
> Ryan: "Oh, my God."
> Todd: "I would just be in constant sorrow for all my days."

Although the desire for young people to control or avoid contact with their families has been documented elsewhere (Ling & Yttri, 2002), the threat of using cellular radios with family members appeared to be even greater than that of other technologies. For example, most participants used mobile phones to speak with their families, while they said they would not use cellular radios.

*Activity awareness*. Participants said they had more information about what the other participants were doing.

> Erica: "I know where everybody is. Like, I usually don't know where Kelly and Dawn are. Like I, I just don't keep up with them that close, to know what they're doing. I don't know like y'all's schedule for work, but now I know when people are working. So much better this week."

This was not only because they spoke more frequently, but also because of the nature of the communication. For example, participants said they often used the cellular radios to ask each other "What's up?" or "How's work?" Participants were positive about this increased awareness.

> Julie: "I actually know what Todd's doing at work today. Cause usually [it's] like 'How was work?' 'Suck. Nothing, nothing happened. I'm so bored.' This is much better."

We note however that we did not get the sense that the participants had the high degree of awareness that has been documented for other systems such as video spaces (Dourish et al., 1996).

*Increased co-present social interaction with participants*. The participants felt that they saw each other more frequently while they had the cellular radios.

> Julie: "I think like I never see Kelly this much. Ever. Like sometimes I get to see her on the weekend."

While they felt the amount they saw each other was affected somewhat by activities that were taking place that particular week, e.g., some of the girls were leasing a new apartment, they clearly articulated that both more frequent talk and increased awareness were key factors in the increased visitations. More frequent talk provided more openings to coordinate co-present activities, as well as being a resource for learning that initiating such activities would be appropriate.

> Todd: "[T]he opportunity to talk with someone more usually leads to like, you know, do you want to go do blah blah blah. And. That just seems kind of a natural thing for me."

Participants were positive about seeing each other more frequently.

## Overall subjective response

Participants said they liked the cellular radios and that they were fun. When asked if they would want to have the service long-term, they were moderately (but certainly not overwhelmingly) enthusiastic. After the study, some of the participants missed their cellular radios and said they were bored without them.

> Julie: "I wanted to talk – I was so bored on my walk home. From class. 'Cause I had no one to talk to. It was really sad."

## Media affordances and their effects

As designers, we mainly wanted to understand how the participants had been able to appropriate the cellular radios for the many communicative activities observed and reported in the study. That is, we wanted to have some idea about which aspects of the technology were critical to this flexibility.

We approached this problem by drawing upon a well-known thread of mediated communication theory, Clark and Brennan's theory of communicative *grounding* (Clark & Brennan, 1991) (henceforth *grounding theory*). Mediated communication theory explores "the relationship between the affordances…of different mediated technologies and the communication that results from using those technologies" (Whittaker, 2003). Grounding theory breaks media affordances into media *constraints* (inherent characteristics of a communication medium), which then in turn affect media *costs* (subjective assessments by users of the appropriateness of specific communicative behaviors). The theory predicts that people both choose media and choose methods of media use in a way that (1) satisfies their communicative purposes and (2) minimizes the "collaborative cost" of communication, i.e., the cost minimization extends across participants and communicative sequences rather than being strictly local. This theory has aspects that could be considered problematic for some purposes – for example, several "axes" of their spaces of media constraints and costs are non-independent, which leads to conflation of certain issues – but we found it to be a useful tool for organizing our thoughts about the issues raised by our data.

In this section, we discuss the affordances of the cellular radios and contrast them with the affordances of several other communication media. We begin by defining the affordances and presenting a summary of the affordances of several technologies. We describe how these affordances relate to higher-level communication phenomena that we observed, and then relate these back to the communicative activities and purposes reported in the previous section.

## Affordances

While we have conducted a more detailed analysis (including other affordances from Clark and Brennan as well as additional affordances we have identified), in the space available here, we will focus on only the most relevant affordances from Clark and Brennan. These are:

*Audible*. The cellular radios transmit voice rather than text.

*Cotemporaneous*. In cotemporaneous media, the listener receives content at approximately the same time the speaker produces it. Cellular radios are cotemporaneous – speech is delivered as it is produced, delayed only by network latency. Email and IM are not cotemporaneous since messages are sent as units.

*(Not) Simultaneous*. With a simultaneous channel, both speakers can send and receive simultaneously. This equates to a full-duplex channel, e.g., telephony. Cellular radios, by contrast, implement a half-duplex channel, in which only one speaker can transmit at a time.

*(Not) Reviewable*. Reviewable media allow transmitted content to be reviewed after receipt at the recipients' leisure. "Reviewable" can be considered a synonym for "browseable" or "persistent." Cellular radio messages are not reviewable; the voice content is delivered at once and can not be replayed. (Note that audible media can be reviewable – consider voice mail.)

*(Low) Production cost*. The incremental cost of producing an utterance varies by media, e.g., people can generally speak more quickly than they type.

*(Low) Reception cost*. Costs are associated with receiving an utterance, e.g., listening costs are typically lower than reading costs (Clark & Brennan, 1991).

*(Low) Start-up cost*. Start-up cost is that of beginning a new discourse. With many media, it involves several components, e.g., getting physical access to the communication device, looking up a number, dialing, etc. Participants were clear that start-up costs were extremely low for cellular radios.

> Dawn: "[T]his was right there, so I didn't have to do any button, dialing thing, plus I don't remember her number."

*(Low) Delay cost*. Costs are associated with delays in replying to an utterance. Some of the most important are social, e.g., delays may be interpreted as rude. Generally, cotemporaneous media have high delay costs – media that are "more like" face-to-face conversation often lead to corresponding behavioral expectations. However, as we discuss further below, even though cellular radio is cotemporaneous, it has low delay costs.

*(Medium) Speaker change cost*. The cost of changing speakers varies depending on the media, e.g., when people are face-to-face, the cost of changing speakers is low because the participants can use resources such as gaze and overlapping speech to facilitate turn-taking. When participants use cellular radios, the cost of changing speakers is somewhat higher, largely because the cellular radios do not support simultaneous speech.

| AFFORDANCES | Email | SMS | IM | Phone | Cellular Radio |
|---|---|---|---|---|---|
| *Audible* | | | | X | X |
| *Cotemporaneous* | | | | X | X |
| *Simultaneous* | | | | X | |
| *Reviewable* | X | X | X | | |
| *Production Cost* | High | High | High | Low | Low |
| *Reception Cost* | High | High | High | Low | Low |
| *Start-up Cost* | High | Low | Medium | High | Low |
| *Delay Cost* | Low | Medium | Low | High | Low |
| *Speaker Change Cost* | High | High | Medium | Low | Medium |

**Table 1. Analysis of communication media used by participants.**

In Table 1, we summarize the affordances of the media commonly used by the study participants. While many discussion points in the remainder of the paper may relate to multiple affordances, we refer only to the most relevant or interesting affordances for particular points, in order to simplify the discussion.

## Basic communication phenomena

Understanding a medium's affordances is mainly useful to the degree that it helps us understand the "conversation that results" from using it. In fact, when we look at the phenomena that arose in our study, we see that tensions between affordances would have made it hard to predict some of the results. In this section, we discuss several phenomena that might be considered counter-intuitive in an audible, cotemporaneous media.

*Reduced feedback*. The audibility and low production cost of cellular radio would suggest that the kind of minimal conversational responses that are normally provided by an attentive listener in face-to-face conversation – often called "backchannel" or "continuers" – would be relatively easy to deliver.

In fact, the non-simultaneous (one-at-a-time) nature of the channel precluded participants from giving feedback while someone else was speaking. That is, while such feedback could be given, it could not be positioned in the natural place relative to (i.e., in overlap with) the other participant's speech. Workarounds were generally unsatisfactory; several participants reported transmitting "fake" (non-spontaneous) laughter after another participant completed a funny utterance. In general, participants produced less feedback, which is known to reduce the fluidity of turn-taking (Whittaker, 2003) and increase speaker change cost.

While this inability to provide natural feedback was occasionally frustrating, participants felt it was also somewhat liberating not to have to respond continuously, e.g., listeners were not under the same obligation to make sympathetic noises in response to a story as they would be in a simultaneous interaction. Hence, the non-simultaneity had the less obvious effect of reducing the perceived reception cost of cellular radio.

*Sustained turn-taking*. As anyone who has used a walkie-talkie for extended periods can attest, sustained conversation takes significant effort in a non-simultaneous medium. This is largely due to the speaker change cost (including the reduced feedback discussed above). Some participants stated that the push-to-talk nature of cellular radios made substantive conversation more difficult.

Nonetheless, participants expressed surprisingly little concern about the non-simultaneity of the channel overall, and it did not seem to be a major inhibitor to substantive conversation (as seen in the previous section). The audio nature of the cellular radios (including low production cost and cotemporaneousness) seemed to be enough to enable this behavior. Recall that media-switching from the cellular radios to other technologies that enabled more fluid turn-taking (such as mobile phones) was not observed or reported.

*Reduced interactional commitment*. Becoming engaged in an undesirably long conversation is often cited as a reason to avoid making phone calls (see, e.g., (Grinter & Eldridge, 2001)). In some ways, a cellular radio interaction is not obviously different from one on the phone. For example, the beginning of an interaction on the cellular radio is nearly as intrusive for the recipient as a phone call (since there is an audible "beep" and the sender's voice is immediately heard on the speaker) and seems just as likely to result in a full "attentional contract" (Nardi et al., 2000) between the participants.

In spite of this, participants had a strong sense that contacting someone on the cellular radio did not represent a commitment to a full-fledged conversation such as might occur on the phone.

> Julie: "Like a phone call is a really big commitment for me. You know, it's like I, I totally plan phone calls… I don't call people to just say like, hey, what are you doing?"

Further, the participants did not feel that individual cellular radio interactions required many formalities.

> Dawn: "[Y]ou don't have to be proper, hello, blah blah blah blah blah, conversation, conversation."

> Erica: "This, there's no hanging up. It's just putting it away."

Both reduced commitment and reduced formality are typical of informal face-to-face communication (Whittaker et al., 1994) and are commonly reported in systems in which users remain continuously connected, such as media spaces (e.g., (Ackerman et al., 1997)) and IM (e.g., (Nardi et al., 2000)).

*Reduced accountability*. The immediate nature of cotemporaneous voice communication often makes recipients feel compelled to answer immediately when addressed. For example, users of the Thunderwire audio space who were known to be at their desks (e.g., those who had just been heard in the space) would explicitly signal inattention if they could not answer promptly (Ackerman et al., 1997).

However, recall that responses to cellular radio transmissions were not necessarily required, particularly immediate responses.

> Kelly: "[I feel like I have to answer if somebody says something to me] but not immediately. I can do it on my own time… if I'm like busy or something like that, and then when I get a chance I'll be like, what did you say, what do you want?"

The participants broadly accepted the type of behavior Kelly describes.

> Kelly: "[T]here can be long pauses and nobody cares and so, phones are just so restrictive and the fact that you have to pay attention so much."

> Erica: "I understand to wait if I'm talking to anybody till they're free and stuff [if they don't answer]… I tried to message Julie earlier but it wasn't working and I figured she was probably at work, busy in a meeting or something."

The cellular radios appear to afford the kind of plausible deniability evident in IM (Nardi et al., 2000). A sender has little or no information about the status of a recipient – the cellular radio itself provides no ongoing awareness information (due to the non-simultaneous nature of the channel), and there is no IM-like supplementary presence mechanism. Hence, the recipient is less visible to the sender and less accountable to respond. Because the media is not persistent, accountability is further reduced: there is no guarantee that messages are received.

*Divided attention (multitasking).* With most audible media (e.g., telephones and audio spaces), pursuing additional activities during a conversation can be problematic. Their simultaneous (full-duplex) nature often means that such activities can be heard by the other participant(s). This signals inattention to the conversation and is often perceived as rude, even when responses are not delayed by the inattention. Their non-reviewable nature means that inattention risks loss of information – the pacing of voice communication requires recipients to hear and mentally buffer the utterances as they arrive.

On balance, the cellular radios, although audible, were seen as compatible with multitasking. Unlike most audible media, they are not simultaneous and do not "leak" information about background activity. Full attention was not required because of the freedom to defer responses and the ability to treat monitoring as a background activity (due to the low reception cost of audio).

> Todd: "[T]he phone is more formal and it… takes all of your attention, and the radio can take all of your attention, but you can kind of also kind of back burner it a bit."

## Media affordances and interaction style

We conclude this section by tying our discussion of media affordances and communication phenomena back to the communicative activities described in the previous section. The intent here is to illustrate how this analysis can help relate media affordances to particular activities and purposes.

All of the three interaction styles described in the related work section – focused conversation, bursty conversation, and intermittent conversation – were seen in our participants' cellular radio use, and all appear to be causally aligned with one or more of the communication phenomena (meaning that the factors that facilitate the given phenomenon also facilitate the corresponding style). Focused conversation is aligned with the phenomenon of *sustained turn-taking* and is exemplified by the communicative activity we termed *substantive conversation*. Bursty conversation, characterized by short turn sequences separated by lapses in talk, is aligned with *reduced interactional commitment* and is exemplified by *chit-chat* and *micro-coordination* (and to a lesser degree, by instances of *play* and *extended remote presence*). Intermittent conversation, characterized by long response delays between individual turns, is aligned with *reduced accountability* and *divided attention* and is exemplified by many instances of *play* and *extended remote presence*. The analyses of the phenomena draw on all of the affordances described in this beginning of this section, so it should be plain that the affordance framework has helped us reason about how the participants were able to apply cellular radios to the various communicative activities. In fact, the discussion in the previous subsection suggests that none of cellular radio's affordances (as listed in Table 1) could be changed significantly without "defacilitating" one or more of the interaction styles.

In addition to enabling us to make sense of the existence of the wide range of styles, media affordances were also helpful in reasoning about our most peculiar study finding – the prevalence of intermittent conversation, which participants indicated was a core use of the cellular radios. This style is closely associated with IM by researchers, and the participants independently perceived the cellular radios as being more closely related to IM than to the telephone.

> Kelly: "I think it's really close to IM. Like I really like it that it's so close because you know, it's one message at a time, it's you know, not commit like, not, you don't have to talk for a long time, you can like leave if you want to or like not answer… [It's more like] IM than the phone."

However, as Table 1 indicates, IM is quite different from cellular radio. We do not know of any other audible media in which intermittent conversation has been reported, but the differences go well beyond that. For example, easy reviewability of messages is often cited as an important enabler for many kinds of loosely-coupled interactions in IM (Isaacs et al., 2002; Nardi et al., 2000).

If we look at Kelly's comment above, we see that she is orienting to a perceived similarity between the strict turn-taking protocol of cellular radio and the line-at-a-time protocol of IM. When read in the context of the analysis in this section, the comment suggests that the non-simultaneous affordance of cellular radio (which, as in IM, imposes a slowed-down, "pacing" effect on conversational turns) reduces overall levels of engagement and thereby reduces commitment, accountability and attentional focus. Non-simultaneity, which is usually considered to have negative effects on spoken conversation, may in this case be a key enabler for IM-like interaction styles in audio.

## Discussion

While this study represents a start at a theoretical understanding of cellular radio, there are ways in which we could increase our understanding of its use. Here, we conducted design fieldwork of a very specific target user population, using the cellular radio service as a rough approximation of our own lightweight audio system to identify emergent issues and phenomena. One could obviously extend the study, increasing both the number of participants and the duration. Similarly, one could examine cellular radio use in work environments, e.g., the mobile service workers (building contractors, field repair workers, etc.) who now form Nextel's main market. Finally, one could study the use of telephony, SMS and cellular radio together since all three media are combined in (e.g.) Nextel handsets (which would provide better comparative use data at the expense of some clarity about the separate media).

That said, the fieldwork and analysis reported here increased our understanding of the issues and tradeoffs involved in designing a lightweight voice communication service. Some of these were theoretical, relating to the affordances, whereas others were pragmatic, relating to use and adoption.

*Theoretical considerations.* We found grounding theory to be useful as a means of highlighting important issues. The affordances identified by the framework of constraints and costs provided plausible explanations of many behaviors that we observed in our data. On the other hand, as a qualitative model, it has limitations, which we have explored to some degree already and now amplify here with two examples.

First, while grounding theory provides reasons why the cellular radios could be used for focused conversation, it does not quite explain why media-switching did not occur after such conversations started. Minimizing effort is one possible explanation, since switching does take effort (of a kind not captured by grounding theory). Another is that the switch to a more "immediate" medium (the phone, for our participants) would imply more of a conversational commitment than might be desired given the casual subject matter and, often, a participant's context (e.g., being on break from waiting tables).

Second, bursty conversation is generally associated with open-channel, continuously-connected media such as media spaces and IM – but cellular radio is neither open-channel nor continuously-connected. It does appear to be sufficiently lightweight so as to suggest, or to provide the illusion of, a space – that is, push-to-talk utterance initiation is "spontaneous enough" to enable this style. For designers, it would be useful to know what "enough" means here.

The upshot is that, while grounding theory helped illuminate the issues around various observed behaviors, we (as designers and practitioners) would not have been able to construct convincing predictions by applying it without study data. The model provides little guidance on how to balance the effects of conflicting affordances. In some cases, certain constraints seem to "trump" other constraints, and it can be difficult to understand why, even when the purpose of communication (the other half of grounding theory) is considered.

A possible reason is that some cases of "trumping" may have less to do with a medium's affordances *per se* than with the user's conceptualization of the medium. Voida *et al.* discuss the tensions caused by affordances of a single medium which have close, previously-established associations with other media and therefore with conflicting conventions of use (Voida et al., 2002). Our data suggests a subtly different kind of conceptual interaction between media. For example, cellular radio is very different from IM, yet the study participants largely thought of it as a kind of "audio IM" – and then used it as such, in spite of media constraints that actually *worked against* this use. This might well have happened if only a few participants had exposure to IM; would these patterns have arisen if none had any exposure to IM?

*Design challenges*. This work suggests a number of issues for designers of future mobile voice communication systems, some of which have already influenced the feature set of our own lightweight audio communication system.

One challenge is to provide means for users to maintain exclusive boundaries of access for certain media. The fact that cellular radio access was truly limited to use within the participants' gelled social group was a key positive factor for them. The problem is to maintain this hierarchy of availability in the face of inevitable pressure for increased access from those who are not, in Julie's words, the "fun people."

A second challenge is to understand more systematically how feature changes affect a system's media affordances and therefore use patterns. For example, consider the "audio chat" implemented by Impromptu (Schmandt, Kim, Lee, Vallejo, & Ackerman, 2002), which records all utterances and supports high-speed browsing of the recordings. This kind of reviewability mechanism may disrupt the plausible deniability currently afforded by the cellular radios.

A third challenge is the support of dynamic conversational groupings in a lightweight way. While the group-connection feature was ultimately rejected by our participants as being too heavyweight, it was widely tried out because of its perceived desirability. We have implemented and described one attempt to provide a more lightweight mechanism (Aoki et al., 2003).